%% file: main.tex
\documentclass{sig-alternate-2013}

\permission{Copyright is held by the International World Wide Web Conference Committee (IW3C2). IW3C2 reserves the right to provide a hyperlink to the author's site if the Material is used in electronic media.}
\conferenceinfo{WWW 2015,}{May 18--22, 2015, Florence, Italy.} 
\copyrightetc{ACM \the\acmcopyr}
\crdata{978-1-4503-3469-3/15/05.\\
http://dx.doi.org/10.1145/2736277.2741102}
\usepackage[numbers,square,sort&compress]{natbib}
\usepackage{flushend}

\usepackage{subfigure}
\usepackage{pdfpages}

\usepackage{color-edits}
\addauthor{as}{red}      
\addauthor{cj}{blue}

\newcommand{\xhdr}[1]{\vspace{2mm} \noindent{\bf #1.}}
\renewcommand{\eqref}[1]{Equation~\ref{#1}}

\usepackage{color}

\newcommand{\jenn}[1]{\textcolor{magenta}{\noindent[JWV: #1]}}

\newcommand{\ignore}[1]{}

\newcommand{\squishlist}{
   \begin{list}{{{\small{$\bullet$}}}}
    { \setlength{\itemsep}{1.5pt}      \setlength{\parsep}{2pt}
      \setlength{\topsep}{2pt}       \setlength{\partopsep}{0pt}
     \setlength{\leftmargin}{2em} \setlength{\labelwidth}{1em}
      \setlength{\labelsep}{0.5em} } }
\newcommand{\squishend}{  \end{list}  }

\newcommand{\base}{{\mathtt{base}}}
\newcommand{\bonus}{{\mathtt{bonus}}}
\newcommand{\stand}{{\mathtt{std}}}
\newcommand{\pbp}{{\mathtt{pbp}}}
\newcommand{\high}{{\mathtt{high}}}
\newcommand{\low}{{\mathtt{low}}}
\newcommand{\prbase}[1]{\Pr(\base | #1)}
\newcommand{\prbonus}[1]{\Pr(\bonus | #1)}
\newcommand{\Ustand}{U_\stand}
\newcommand{\Upbp}{U_\pbp}

\begin{document}


\title{Incentivizing High Quality Crowdwork}
%
%

\numberofauthors{2} 
\author{
\alignauthor Chien-Ju Ho\thanks{Much of this research was conducted while Ho was visiting Microsoft Research.}\\
	\affaddr{University of California, Los Angeles}\\
	\affaddr{cjho@ucla.edu} \\
\alignauthor Aleksandrs Slivkins\\
	\affaddr{Microsoft Research}\\
	\affaddr{slivkins@microsoft.com} \\
\and
\alignauthor Siddharth Suri\\
	\affaddr{Microsoft Research}\\
	\affaddr{suri@microsoft.com}\\
\alignauthor {Jennifer Wortman Vaughan}\\
	\affaddr{Microsoft Research}\\
	\affaddr{jenn@microsoft.com}\\
}

\maketitle

\begin{abstract}
\input{abstract}

\end{abstract}


\section{Introduction}
\label{sec:intro}
\input{intro}

\xhdr{Related Work}
\input{related}

\section{Preliminaries}
\label{sec:preliminary}
\input{preliminary}

\section{Does PBP Work?}
\label{sec:whether}
\input{exp1}

\section{When Does PBP Work?}
\label{sec:how}
\input{exp2}

\section{Why Does PBP Work?}
\label{sec:why}
\input{exp3}

\section{Where Does PBP Work?}
\label{sec:generalizability}
\input{exp4}

\section{A Theory of Worker Incentives}
\label{sec:model}
\input{model}

\section{Conclusions and Discussion}
\input{conclusion}

\section*{Acknowledgements}
\input{ack}

\raggedright
\bibliographystyle{abbrvnat}
\bibliography{qcp,dynamicPA,bib-bandits,bib-abbrv-short}
\end{document}

%% file: abstract.tex
We study the causal effects of financial incentives on the quality of crowdwork. We focus on \emph{performance-based payments} (PBPs), bonus payments awarded to workers for producing high quality work. We design and run randomized behavioral experiments on the popular crowdsourcing platform Amazon Mechanical Turk with the goal of understanding \emph{when}, \emph{where}, and \emph{why} PBPs help, identifying properties of the payment, payment structure, and the task itself that make them most effective. We provide examples of tasks for which PBPs do improve quality. For such tasks, the effectiveness of PBPs is not too sensitive to the threshold for quality required to receive the bonus, while the magnitude of the bonus must be large enough to make the reward salient.  We also present examples of tasks for which PBPs do not improve quality.  Our results suggest that for PBPs to improve quality, the task must be \emph{effort-responsive}: the task must allow workers to produce higher quality work by exerting more effort.  We also give a simple method to determine if a task is effort-responsive \emph{a priori}.  Furthermore, our experiments suggest that all payments on Mechanical Turk are, to some degree, \emph{implicitly} performance-based in that workers believe their work may be rejected if their performance is sufficiently poor. Finally, we propose a new model of worker behavior that extends the standard principal-agent model from economics to include a worker's subjective beliefs about his likelihood of being paid, and show that the predictions of this model are in line with our experimental findings. This model may be useful as a 
foundation for theoretical studies of incentives in crowdsourcing markets.

%% file: intro.tex
Crowdsourcing markets are platforms on which workers around the world perform tasks for pay. In a crowdsourcing market like Amazon Mechanical Turk, requesters post tasks along with the amount of money that they are willing to pay workers for their task's completion.  Workers can then browse the available tasks and choose tasks to work on.

Crowdsourcing markets are used to conduct user studies~\cite{KCS08}, run behavioral experiments~\cite{HRZ11,MS12}, collect data~\cite{HC10,WB+11}, test or even build business applications~\cite{Shall12-book,Alonso13-perspective}, and more.  While these markets are effective at recruiting diverse labor pools, the quality of work produced varies widely across tasks and workers.  The prevalence of low quality crowdwork has inspired a growing literature on techniques to boost accuracy, for example, by using redundant assignments for labeling tasks~\cite{SPI08,IPW10,KOS11,LPI12,HJV13}, smartly assigning tasks to workers~\cite{HV12,HJV13}, introducing social incentives~\cite{RKK+11,SHC11}, or altering financial incentives~\cite{MW09,RKK+11,BKG11,SHC11,H11,YCS13,YCS14,GLM14}. These solutions have had mixed success, and how to improve the quality of work in general is still not well understood.

In this paper, we study the use of financial incentives to encourage
high quality crowdwork on Amazon Mechanical Turk.  In particular, we
focus on the use of \emph{performance-based payments} (PBPs), bonus
payments awarded to workers for producing high quality work.  Previous
empirical studies of performance-based payments in crowdsourcing
markets have produced mixed and somewhat contradictory
recommendations.  \citet{H11} and \citet{YCS14} suggested that PBPs
can improve work quality, while \citet{SHC11} found no improvement and
\citet{YCS13} found no difference in quality when varying bonus size.

Our results explain these disparities in prior work.  Furthermore, we show how to
generalize previous findings beyond the particular tasks that were
studied.
We design and run experiments with the goal of understanding
not just whether PBPs improve work quality for a specific task or
bonus size, but \emph{when}, \emph{why}, and \emph{where} they improve work quality.
We identify properties of the payment, payment
structure, and the task itself that make PBPs effective.

\ignore{
In this paper, we seek to understand what causes this disparity in
results and how to generalize previous findings beyond the particular
tasks that were studied.  We are motivated by the question of what it
is that determines whether PBPs improve performance for a particular
crowdsourcing task, and how the parameters of the payment scheme used
(size of the bonus payments, rules for determining whether or not a
bonus is received, etc.) effect work quality.
}

In our experiments, we first identified a task (proofreading an article)
for which PBPs improve workers' performance.  We tested the robustness of
this finding by varying the payment structure and amount.  We found that
using PBPs with a wide range of quality thresholds improved work quality provided
the bonus awarded for exceeding the threshold was sufficiently high.  We
also found that even when standard, unconditional payments are used and no
explicit acceptance criteria is specified, workers may behave as if the
payments are \emph{implicitly} performance-based since they believe their
work may be rejected if its quality is sufficiently low. 

We examined potential reasons why PBPs improve quality. We found that simply increasing the amount of the base payment without offering any bonus significantly improved quality, contradicting several previous studies~\cite{MW09,RKK+11, BKG11, Araujo13, GLM14}.  However, PBPs led to improved quality and lower cost compared to a guaranteed payment of the same amount. We also found that whether the opportunity to receive a bonus or higher base payment is revealed before or after the task is accepted does not make a difference in the quality of crowdwork, ruling out the possibility that the increased quality we observed was due (at least in part) to the reciprocity caused by workers' joy at receiving an unexpected bonus, as discussed by \citet{GLM14}.

Finally, we investigated which properties of a particular \emph{task} allow PBPs to have an 
effect.  We examined the conjecture that  PBPs are more likely to improve quality 
on \emph{effort-responsive} tasks, tasks for which workers can produce 
higher quality work by exerting additional effort.  
We ran experiments on four different tasks.
By taking the amount of time workers spent on the task as the proxy measure for 
their effort, we found that additional effort was correlated with improved 
quality for the tasks for which PBPs helped, but not for the tasks for which PBPs did not 
help.  This observation yields a simple method for requesters to determine whether 
or not a given task is likely to benefit from PBPs.

Based on our experimental results, we propose a simple theoretical model of worker behavior. The model is a variant of the standard principal-agent model from economics that additionally incorporates workers' subjective beliefs about the quality of work required to be paid.  We show that this model can be used to explain our key empirical observations which cannot be explained using the principal-agent model alone.  This model may be useful as a more realistic foundation for future theoretical work on crowdsourcing markets.


%% file: related.tex
Performance-based payments have been studied extensively outside of
crowdsourcing markets. \citet{Lazear00} conducted a highly influential
analysis of observational data from an autoglass company that switched from
paying workers a fixed hourly rate to paying workers based on the number of
units installed, and showed that workers' performance significantly
improved when payments were contingent on work done. As another
example, \citet{GR00} gave college students fifty questions from IQ tests
to answer and found that their performance increased when they received a
bonus for answering questions correctly as long as the bonus was
sufficiently high. On the other hand, when the bonus payment was very
small, quality decreased compared with not offering a bonus at all.
The
psychology literature suggests that this is perhaps due to a decrease in
workers' intrinsic motivation (enjoyment, responsibility, pride) for
performing well, though this theory is not universally
accepted~\cite{FO97,CP94,EC96}.

\citet{CH99} performed a meta-analysis of 74 papers examining the effect of
using payments contingent on performance in lab experiments.  They showed
that using payments contingent on performance improved the average
performance of subjects for tasks in which increased effort leads to
improved performance, such as memory or recall tasks, clerical tasks such
as coding words or building things, and problem-solving tasks.
Similar
meta-analyses were performed by \citet{J+98}, \citet{B+00},
and \citet{HO01}.
We build on and extend this work by showing that PBPs have a causal impact
on work quality in a field setting, specifically, a crowdsourcing
environment in which eliciting high quality work is a main concern for requesters.  Moreover, we give evidence to support
the conjecture of \citet{CH99} that PBPs work in tasks that are effort-responsive.


The results of studies on performance-based payments in crowdsourcing
markets have been mixed and sometimes discouraging.  On the positive
side, in an early workshop paper, \citet{H11} showed that when asking
workers to evaluate the relevance of resumes, PBPs increased both
performance and the time workers spent on the task.  In very recent work, \citet{YCS14} studied a setting in which workers switched back and forth between two types of tasks, and showed that PBPs improved performance, especially when used immediately after a task switch.

On the negative side, \citet{SHC11} compared fourteen different
incentive schemes, including four using PBPs,\footnote{The differences
in these four treatments were whether payments were described in terms
of rewards for high quality or punishments for low quality, and
whether quality was measured objectively or in terms of agreement with
other workers' responses. Of these, the only treatment that resulted
in performance statistically significantly different than the control
was punishing workers when their responses did not agree with others.}
and saw little variation in their effects on quality of work.
However, the bonuses offered were extremely small, only 10\% (\$0.03)
on base payments of \$0.30. One might hypothesize that the lack of
effect stems from decreased intrinsic motivation as in \citet{GR00},
or that obtaining the small bonus is simply not worth the costly
additional effort that would be required.  More recently,
\citet{YCS13} studied the effect of varying the bonus size of PBPs and
found that the size of the bonus did not impact the quality of work.
In their case, all bonuses offered were large compared with the base
payment (\$0.04, \$0.08, \$0.16, or \$0.32 on a base payment of \$0.01
per task), so it is possible that even their smallest bonus was large
enough to elicit the workers' maximum effort; since \citeauthor{YCS13}
did not include a treatment without PBPs, there is no way to know
whether PBPs boosted quality compared with offering base payments
alone.  In Section~\ref{sec:how}, we give a unifying explanation for
these results.  We show small bonuses in a PBP result in little to no
effect on work quality, and that it can be hard to detect the effect
between two large PBP bonuses.  However, the overall trend is that
using PBPs with sufficiently high bonuses yields better quality work.

Several additional studies have examined financial incentives in
crowdsourcing markets. \citet{HC10} empirically estimated workers'
reservation wages and found that many workers aim to hit payment
targets (such as multiples of \$0.05).  
\citet{MW09} found that paying
workers more increased the number of tasks workers chose to complete,
but did not increase performance on each task. That increased pay did not
increase performance was observed by other authors as
well~\cite{RKK+11, BKG11, GLM14, LRR14}.  In this paper, we 
found that paying workers more can actually increase their performance for some
types of tasks. This disparity can be explained by considering 
workers' subjective beliefs on how much work they must do to get their work 
accepted. In prior work, workers either already performed well even with low pay 
since the tasks were easy or were given additional instructions which
could have primed their
subjective beliefs.
In our experiments, workers are uncertain about how much work they should do
to get paid. Therefore, they are willing to produce higher quality work to
increase their chance of having their work accepted when the payments are higher.

\ignore{
\citet{MW09} found that paying
workers more increased the number of tasks workers chose to complete,
but did not increase performance on each task; it is possible that
this is due to a priming effect from the tutorial they presented to
workers, which may have altered workers' subjective beliefs about what
they must do to have their work accepted.  That increased pay did not
increase performance was observed by other authors as
well~\cite{RKK+11, BKG11, Araujo13, GLM14, LRR14}.  One explanation
could be that their tasks were not effort-responsive, a concept we
explore in Section~\ref{sec:generalizability}. \jenn{I edited this
paragraph but am not happy with it. CJ, thoughts? Should we claim less
and point the reader forward as we did before?}
} 

\citet{GLM14} showed that the manner in which the payment is
presented can influence quality.  In particular, they found that on oDesk, 
a crowdsourcing market for larger tasks, initially telling workers they
would receive \$$3$ per hour and increasing this payment to \$4 after the
job was accepted led to higher performance than paying either \$$3$
or \$$4$ per hour without the surprise.  We show in Section~\ref{sec:why}
that this result did not translate to the Mechanical Turk setting.

\citet{HSV14} studied the algorithmic problem of adaptively optimizing PBPs in crowdsourcing markets, modeling the problem as a dynamic variant of the standard principal-agent model from contract theory.  Their model assumed that each worker chooses how much effort to exert in order to maximize his expected utility, which is simply his expected payment minus the cost of his effort.  In Section~\ref{sec:model}, we propose a variant of this worker model that is in line with our experimental observations. The results of \citet{HSV14} still apply when our worker model is used in place of theirs.

%% file: preliminary.tex
Before describing our experiments, we briefly describe the setting in which
they were conducted.  All of our experiments were run on Amazon Mechanical
Turk\footnote{https://mturk.com}
(henceforth MTurk), one of the most
popular crowdsourcing platforms. On MTurk, requesters post tasks (HITs) for
workers to complete.  When a worker browses a task, he sees a description
of the work to be done along with the amount of money that the requester
has offered as a base payment.  A worker can then choose whether to accept the
task.  After the worker completes the chosen task, the requester may
evaluate the worker's submission and choose to either approve or reject the
work.  Each worker has an approval rating which is simply the fraction of
HITs he has submitted that have been accepted.  If the work
is rejected, the worker is not paid and his approval rating, which serves
as a \emph{de facto} reputation score, suffers. If the work is accepted, the
worker receives the base payment for the task and his approval rating
increases. Requesters also decide at this time whether or not to award a
bonus payment on top of the base, and how much of a bonus to award.  The
possibility of such a bonus may or may not be included in the task
description.

HITs can also have qualifications associated with them.  Only those workers
who have the appropriate qualifications can do a HIT.  These qualifications
are specified by the requester.  For example, we used a geographic
qualification to restrict our tasks to workers located in the United
States, and used qualifications to disallow workers from completing
the same type of task more than once. We also used qualifications for
random assignment in one of our experiments; how and why we did this is described in
Section~\ref{sec:why}.

Our experiments focus on the use of threshold-based PBPs.  A
threshold-based PBP is specified by a base payment, a bonus payment, and a
threshold.  The base payment specifies the amount of money a worker
receives from completing the task; this is fixed at \$$0.50$ USD in all of
our experiments.\footnote{The base payment of $\$0.50$ was chosen so that
workers could obtain a \$$6$ hourly rate from the base payment alone with a
reasonable amount of effort in each of our tasks.}  The bonus payment specifies the amount the worker can potentially receive as a bonus.  The threshold determines what the worker must do in order to obtain the bonus.

All of
the experiments in this paper were approved by the Microsoft Research IRB.

%% file: exp1.tex
Our first experiment was designed with two goals in mind. The first
was to verify that PBPs can lead to higher quality crowdwork and
identify a task for which this happens. The second was to determine if
there exists what we call an {\em implicit PBP effect}: even if the
requester offers a guaranteed payment, MTurk workers have
subjective beliefs on the quality of work they must produce in order
to receive this payment, and therefore behave as if the payments were
(implicitly) performance-based.
We measure these subjective beliefs by the difference in 
the quality of crowdwork when base payments are
explicitly guaranteed, effectively resetting the workers' subjective
beliefs, compared to when base payments are not guaranteed.

\subsection{Experiment Design}
\label{sec:pr}

In this experiment, workers were asked to proofread an article of 500
to 700 words and correct spelling errors. For each article, we
randomly inserted 20 typos from a list of common spelling errors.
Workers were asked to input the line number of each typo, the
misspelled word, and the correct spelling of the word.

This task has two key properties. First, we would
expect that workers could produce better work by exerting more
effort---the more carefully a worker reads or the more passes a worker
takes over the text, the more typos he will find---and that this would
open up the possibility of PBPs improving quality.  (We study this
conjecture in more detail in Section~\ref{sec:generalizability}.)
Second, since we injected the typos into the text, the quality of each
worker's output could be measured objectively, though this was not
known to the workers.

Before accepting the HIT, each worker saw a
preview consisting of the instructions, an example article with typos,
and the base payment of \$$0.50$ USD. The preview was the same for all
workers.
After workers accepted the HIT, they were randomly assigned to
different treatments and then shown treatment-specific instructions,
when applicable. Our experiment had a $2 \times 3$ design, with
$2$ treatments governing the base payment and $3$
treatments governing the bonus payment (if any). We discuss the bonus
treatments first:

\squishlist
	\item \emph{No Bonus:} This is the control group. It had no bonus and
  	no mention of a bonus.
  	\item \emph{Bonus for All:} All workers earned a \$$1$ bonus after submitting the HIT.
  	\item \emph{PBP:} Workers earned a \$$1$ bonus if they found $75$\% of the typos found by the other workers.
\squishend

The purpose of the control group was to allow us to measure the baseline
number of typos workers found.  The purpose of the Bonus for All treatment was
to test if simply paying more resulted in higher quality work.  The purpose
of the PBP bonus treatment was to test if specifically incentivizing for
quality improved the number of typos found.


As mentioned, we are also interested in whether workers have
subjective assumptions on how much effort they must exert to get
their work accepted. Workers may be afraid that if they do not find a
sufficient number of typos their work will be rejected, resulting in
no pay and a negatively affected MTurk reputation. To estimate this,
we designed a treatment in which workers were explicitly guaranteed
acceptance provided that they completed a very small amount of work.
We had two treatments for the base payment:


\squishlist
	\item \emph{Non-Guaranteed:} There were no extra
	instructions.  This is the control and emulates most MTurk tasks.
      \item \emph{Guaranteed:} Workers were told they would get
        paid if they found at least one typo.
\squishend

The first typo appeared before line $3$ in each article.  Thus a
worker would only have to do a trivial amount of work to ensure they
got paid in the guaranteed base treatment.


\subsection{Results}
\label{sec:exp1results}

\begin{figure}[h]
   \centering
   \includegraphics[width=0.8\columnwidth, keepaspectratio]{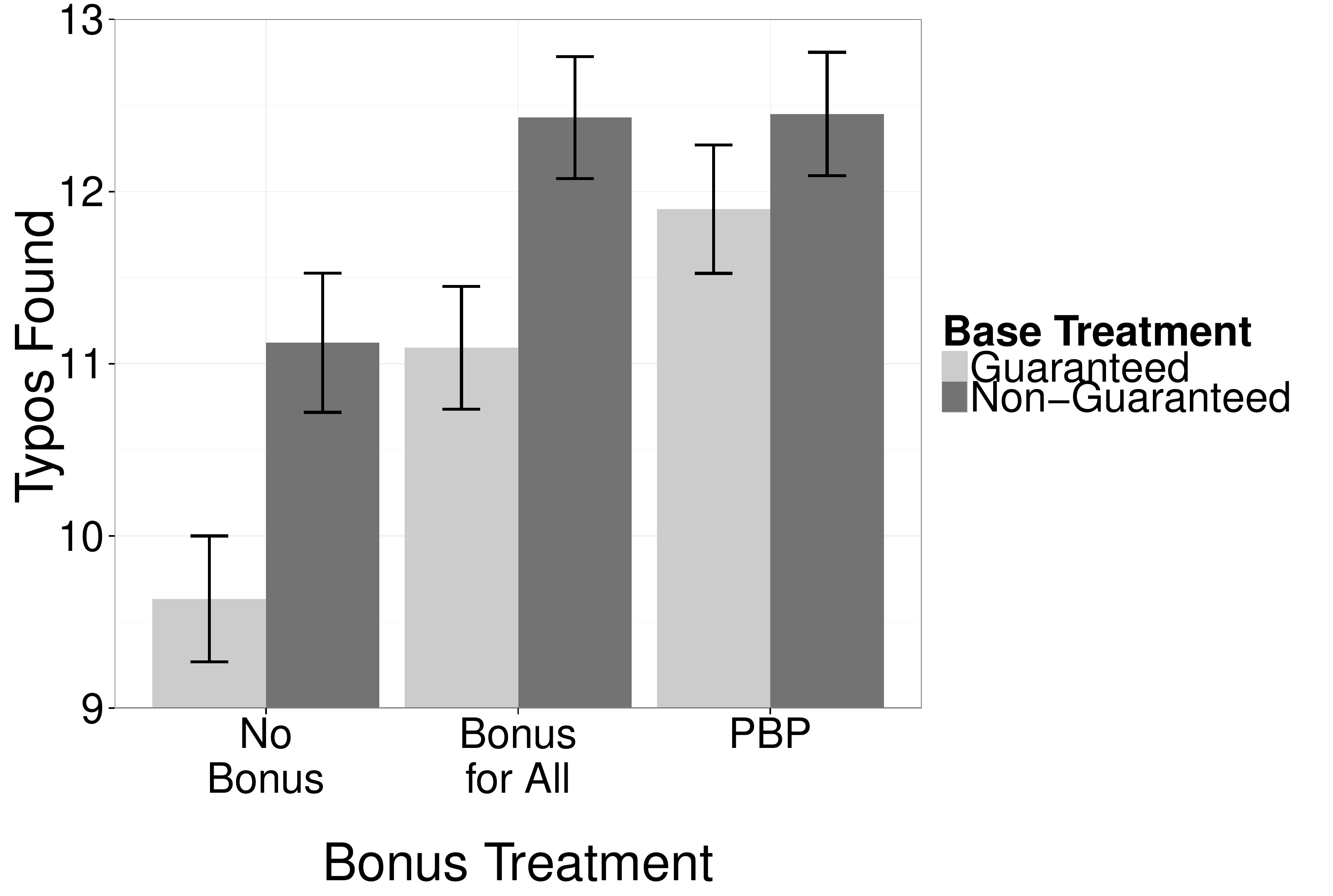}
   \caption{The effect of different payment
   schemes on work quality in the proofreading task. Error bars indicate
   the mean $\pm$ one standard error.}
   \label{fig:qcp_work}
\end{figure}

The HIT was completed by 1,000 unique workers, who were each assigned uniformly to one of the six treatments.
\ignore{
We removed the results
from 1 worker since we received an email from this worker indicating
that she/he accidentally submitted the HIT without doing any work.
}
We conducted a chi-squared test to check for significant differences
in the number of participants who finished the six treatments and
found none ($p = 0.38$).  The primary dependent variable (or outcome
measure) we were interested in is the number of true typos found.
In the analysis we made six comparisons that we spell out below. We
performed this analysis using an ANOVA with one-sided, planned comparisons \cite{seltman-book} and report p-values that have
been corrected for these multiple (six) comparisons. The results of
this experiment are shown in Figure~\ref{fig:qcp_work} and described
below.


\xhdr{PBPs improve quality} To determine whether PBPs increase
quality for this task, we focus on the non-guaranteed base treatments
since almost all HITs on MTurk do not explicitly guarantee any kind of acceptance criteria. Workers in the PBP bonus treatment found on average 1.3 more typos than workers in the No Bonus treatment ($p=0.042$), showing that PBPs did improve quality for this task.

\xhdr{All payment schemes may be implicitly performance-based} In the No Bonus treatment, the
guaranteed base resulted in 1.5 fewer typos found on average
compared with the non-guaranteed base ($p = 0.015$). Similarly, in the Bonus for All treatment, the guaranteed base resulted in 1.3 fewer typos found on average ($p = 0.024$). While there may be other explanations, this suggests that workers do have subjective beliefs on the amount of work that needs to be done for their work to be accepted, lending support to our conjecture that payments on MTurk are already implicitly performance-based. We discuss this further in Section~\ref{sec:model}.

In the PBP bonus treatment, we did not see a significantly
different effect between the guaranteed base and non-guaranteed base
treatments.
We offer two related explanations of this finding.  First, the only
way to grant a bonus using the MTurk API is to first accept the work.
This means that in the PBP bonus treatment, workers would likely
believe that finding 75\% of typos would almost certainly result in
their work being accepted, already altering their subjective beliefs.
Second, the treatment might have made this 75\% threshold more salient
to the workers.  This gave a clear goal for the workers to strive for.

\xhdr{Simply paying more improves quality} Focusing again on the
non-guaranteed base treatment, workers in the Bonus for All treatment
found on average 1.3 more typos than workers in the No Bonus treatment
($p=0.036$). Thus offering an unconditional bonus---which is
essentially just paying more---increased quality.

This finding is perhaps surprising since it appears to contradict the
results of prior work~\cite{MW09,RKK+11,BKG11}. We give two potential
explanations. First, since the announcement of the bonus came after
workers accepted the HIT, the workers may be exhibiting reciprocity by
doing higher quality work~\cite{GLM14}, rewarding the requester for
this pleasant surprise.  We further test and refute this hypothesis in
Section~\ref{sec:why}. Second, this could be explained by the implicit
PBP effect described above. That is, workers might have subjective
beliefs about the number of typos they must find to get paid. If we
increase the bonus payment, workers might be willing to put in more
effort to increase their probability of earning this higher amount.

This observation is not inconsistent with previous work. In most prior
work, either easy tasks were chosen which might cause workers to
perform well even for low pay~\cite{RKK+11, BKG11} or additional
instructions or tutorials were provided which may have primed workers'
subjective beliefs~\cite{MW09}.

\xhdr{PBPs can save money compared with high unconditional payments}
In the non-guaranteed base treatment, the difference in the number
of typos found in the PBP and Bonus for All treatments is not
significant. Both resulted in higher quality work than the
control. However, we spent much less money on the PBP
treatment. We paid each worker $\$1.50$ in the Bonus for All
treatment, while we paid each worker only $\$0.97$ on average in the
PBP treatment with non-guaranteed base and $\$0.96$ on average
in the PBP treatment with guaranteed base.  Therefore, it may
still be advantageous for requesters to offer PBPs even if they could
achieve the same quality work with unconditional payments.

\ignore{
\xhdr{Summary}
These are the conclusions from our experiment on the proofreading task.
\begin{itemize}
	\item PBPs work: they increase workers' performance.
	\item The unconditional bonus also performs better than the control.
	\item Explicitly guaranteeing the base payment improves performance,
suggesting that payments on MTurk are already implicitly performance-based.
\end{itemize}
}

%% file: exp2.tex
Having established that PBPs can improve quality for the proofreading
task, we investigated the effect of varying two parameters of the
payment scheme, the bonus threshold and the bonus amount, to better
understand when PBPs help.

\subsection{Bonus Thresholds: Experiment Design}
\label{sec:threshold}

We first tested the effect of varying the threshold of quality that
must be met in order for workers to receive the bonus.  We used the
same proofreading task described in Section~\ref{sec:pr}, with the
same base payment of $\$0.50$ and bonus of \$$1$. Workers were
randomly assigned to treatments in which they were told they could
earn the bonus if they found at least 5 typos or at least 25\%, 75\%,
or 100\% of the typos found by the other workers. In the control,
workers did not receive any bonus or see any mention of a bonus.


\subsection{Bonus Thresholds: Results}
\begin{figure}[h]
   \centering
   \includegraphics[width=0.8\columnwidth, keepaspectratio]{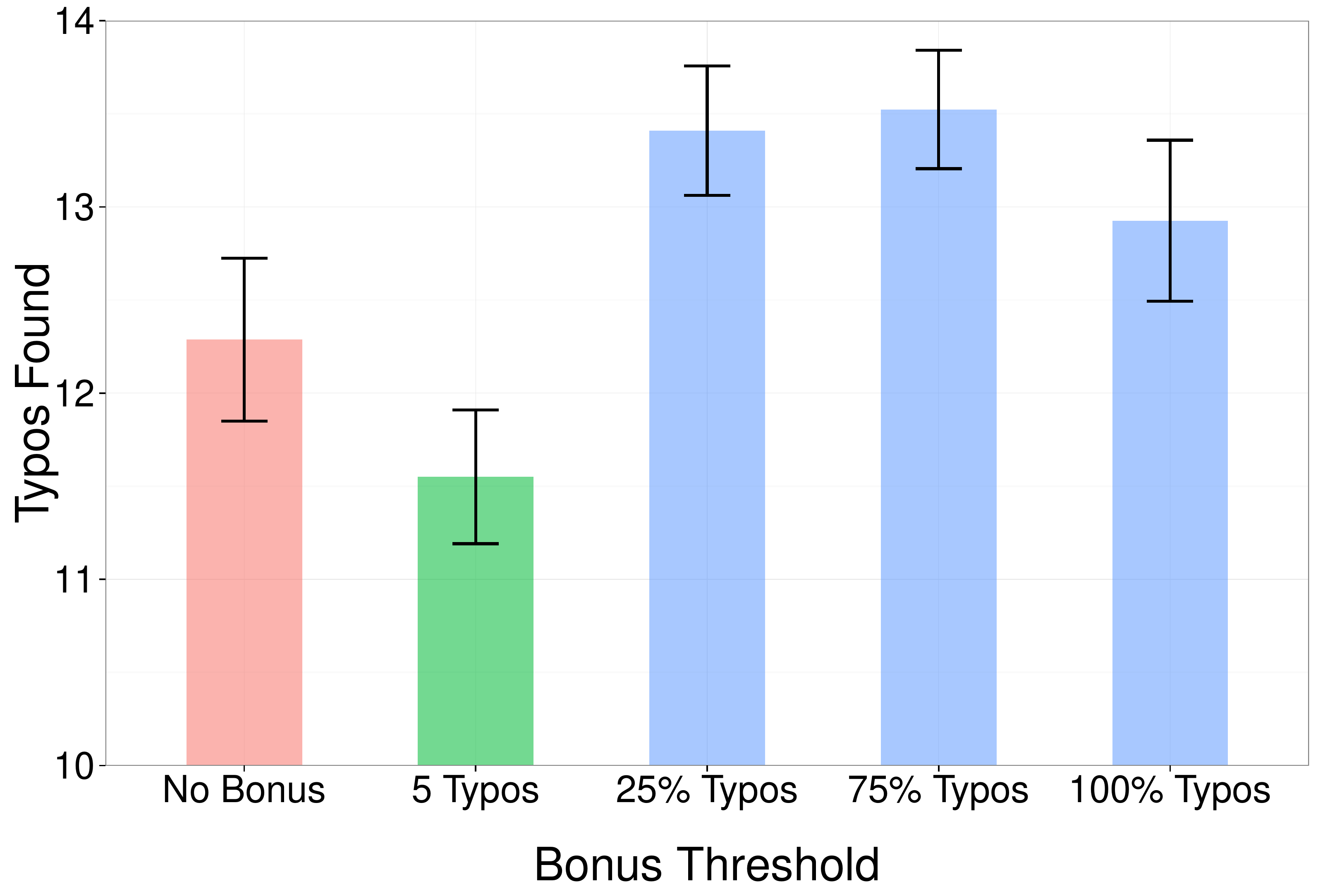}
   \caption{The effect of the bonus threshold on work quality in the
     proofreading task. Error bars indicate the mean $\pm$ one standard error.}
   \label{fig:bonus_threshold}
\end{figure}

%
%
The results from 585 unique workers are presented in Figure~\ref{fig:bonus_threshold}. As
before, we ran an ANOVA with one-sided, planned comparisons \cite{seltman-book} and
report p-values that have been corrected for the multiple (five)
comparisons we describe below.

\xhdr{PBPs improve quality for a wide range of bonus thresholds}
Improvements in quality over the
control can be observed in both the 25\% and 75\%
 treatments.
The 25\% treatment resulted in workers finding, on average, 1.1 more
typos than the control
 $(p=0.082)$.  Similarly, the 75\% treatment resulted in workers finding, on
 average 1.2 more typos, than the control $(p=0.049)$. 
We conjecture that setting a threshold
anywhere between 25\% and 75\% would yield similar results, so the
improvements from PBPs are not overly sensitive to the threshold.
However, the 100\% typo condition was neither significantly different than
the 75\% treatment or control, so 
our results suggest that if the bonus threshold is set too
high, then workers' average performance slightly
decreases. This could be due to some workers giving up because they do
not believe the bonus is attainable.

\xhdr{Subjective beliefs on the quality thresholds can improve work}
Since each article contains
20 typos in total, 25\% of the typos found by other workers is at most
5 typos and will be exactly 5 typos if the number of workers is
sufficiently large.  In this sense, the two thresholds are roughly
equivalent.
However, workers in the 25\% treatment performed much
better than workers in the 5 typos treatment.
Workers in the 25\% treatment found, on average, 1.9 more typos than those
in the 5 typo treatment ($p<.001$).
An analysis of the line numbers in which typos were found showed that
workers in the  5 typo  treatment did not stop reading before workers in
the 25\% treatment; however, there may be a variety of
other explanations for this.
For example, it is possible that 
workers in the 25\% group had different subjective beliefs and
thought that they would need to find more than 5 typos to receive the bonus.
After all, workers did not know the total number of typos in the article.

\subsection{Bonus Amounts: Experiment Design}
\label{sec:amount}

We next examined the effect of varying the bonus amount.  We used
the same proofreading task with a base payment of \$$0.50$.  Workers
were assigned to treatments in which they could earn either \$$0.05$,
\$$0.50$, or \$$1$ if they found $75\%$ of the typos found by other
workers.  Once again, workers in the control did not receive 
or see any mention of a bonus. 

As an implementation detail, we note that these experiments were run
simultaneously with those described in Section~\ref{sec:threshold},
allowing us to share two treatments (the control and the \$$1$ for
75\%) and run only seven treatments in total instead of nine.  We
excluded workers who had already participated in the experiment from
Section~\ref{sec:whether}.
We collected results from 815 unique workers assigned uniformly
to the seven treatments.
A chi-squared test showed no significant differences
between the number of workers who completed the seven treatments ($p =
0.23$)

\ignore{
Workers were
randomly assigned to one of the treatments in one of the
experiments. This allowed us to share two treatments between these
experiments: the control condition and they \$$1$ bonus for finding
75\% of the typos. We collected data from 815 workers in
total. Finally, workers who participated in the experiment in
Section~\ref{sec:whether} were not allowed to do either the bonus
amount or bonus threshold experiments.
}

\subsection{Bonus Amounts: Results}
\label{sec:amountresults}

\begin{figure}[h]
   \centering
   \includegraphics[width=0.8\columnwidth, keepaspectratio]{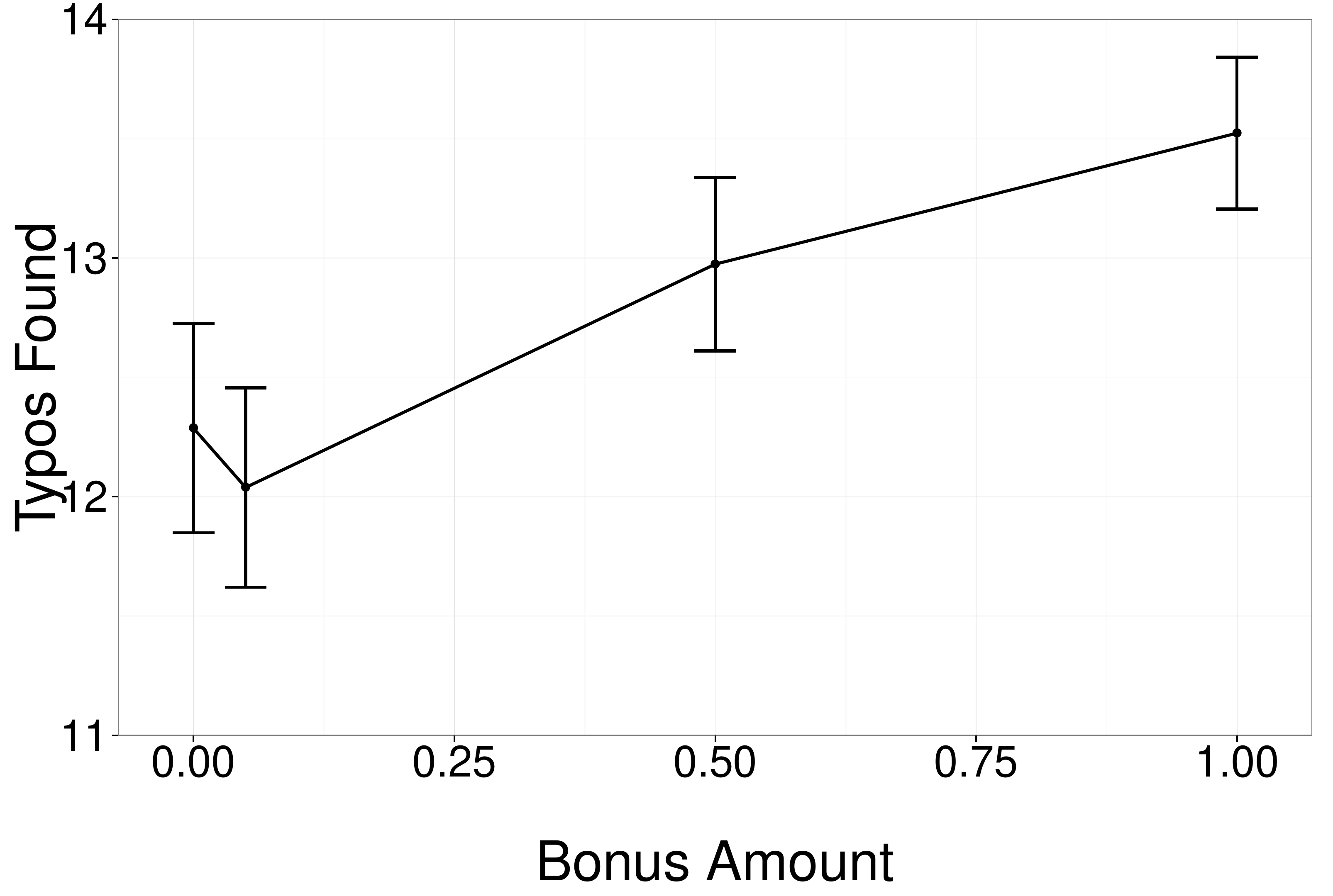}
   \caption{The effect of the bonus amount on work quality in the
     proofreading task. Error bars indicate the mean $\pm$ one standard error.}
   \label{fig:bonus_amount}
\end{figure}

We collected data from 451 unique workers.  
Figure~\ref{fig:bonus_amount}
shows the overall trend:
PBPs lead to higher quality work
only when the bonus is sufficiently large, but increasing the bonus
amount has diminishing returns.
Indeed, regressing the number of typos found
on the bonus amount shows that an extra \$1 of bonus results in
finding 1.4 more typos $(p = 0.002)$ on average.

This may help to explain previous negative results on PBPs in
crowdsourcing markets.  \citet{SHC11} reported little or no quality
improvement using PBPs compared with fixed payments.  However, they
offered a bonus payment of only \$$0.03$, 10\% of their \$$0.30$ base.
As we observe from the leftmost two points in
Figure~\ref{fig:bonus_amount}, PBPs do not improve quality when the
bonus is very small.  \citet{YCS13} tested PBPs with several bonus
sizes and reported that bonus size alone did not significantly impact
quality.  However, their study did not include a control with fixed
payments, and the bonus sizes that they tested were all significantly
larger than their base payment (\$0.04, \$0.08, \$0.16, and \$0.32 on
a base payment of \$0.01).  As we observe from the rightmost two
points in Figure~\ref{fig:bonus_amount}, increasing the bonus size
only leads to minor improvements in quality once the bonus is already
sufficiently large.  It is only when we view the whole picture that we
can see that PBPs do help. 

\ignore{
 In particular, when the bonus is small
(\$$0.05$, or 10\% of the base payment), the effect of
performance-based payments is hard to detect.  \citet{SHC11}, who
reported little or no quality improvement using PBPs, also offered a
10\% of their base payment as a bonus.  Additionally, when the bonus
is large enough (\$$0.50$ and \$$1$), there are no significant
differences between the effects found using different bonus
amounts.  This is in line with ~\citet{YCS13}, who tested several
bonus sizes considerably larger than their base payment and reported
that bonus size alone did not significantly impact quality.
}

\ignore{
The result suggests that the effect of PBP can be observed in a wide
range of bonus amounts as long as the bonus is large enough,
}

Taken as a whole, the results in this section show that PBPs improve
quality for a wide range of possible thresholds,
provided that the requester offers a bonus that is high enough to make
the extra reward salient.


%% file: exp3.tex
There are two primary motivations for our next experiment.  First, we
wanted to verify that PBPs are useful in other tasks beyond finding
typos.  Second, we wanted to explore potential reasons why PBPs work.
In particular, as pointed out in Section~\ref{sec:exp1results}, simply
increasing the amount of the bonus payment led to almost as much of an
improvement as using PBPs in the proofreading experiment.  While it
could be that workers are responding rationally to the
provided incentives, it could also be the case that workers are
increasing their effort due to a reciprocity effect; workers are
pleasantly surprised to discover the opportunity to receive a
(performance-based or unconditional) bonus after accepting the HIT,
and reward the requester for this kind action by working harder.
Indeed, \citet{GLM14} found, in a different crowdsourcing context, that
workers who accept a task and then receive an unexpected bonus do
higher quality work than workers who are paid the same amount total
but are told up front.  This experiment is designed to test whether
this ``unexpected bonus effect,'' is the (partial)
cause of the observed increases in performance using PBPs.

\ignore{
For example, are
workers exerting more effort to increase their probability of getting
the higher bonus?  Alternatively, could the increased effort be due to
a reciprocity effect?  \citet{GLM14} found, in a slightly different
context, that workers who accept a task and then receive an unexpected
bonus do higher quality work.  So we ask, are workers rewarding the
requester for his/her kind action of paying more by their own kind
action of putting in more effort?
}


\subsection{Experiment Design}

In this task, workers were shown twenty pairs of images. Ten of the
pairs were identical images, while the other ten pairs contained minor
differences.  Workers were asked to specify whether each pair was
identical or not, and were not told how many pairs of images were
identical in advance.  Again, this task has two key properties we
desire. First, we speculated that workers would be more likely to spot
the differences between images if they spent more time and effort
looking. Second, we can objectively measure the quality of workers'
output by the number of correctly answered pairs.  A similar task was
used in experiments by \citet{YCS13}.

To test our research questions, we wished to vary the bonus amount and
bonus rules as well as the amount of the base payment.  Obviously, if
we launched two HITs that differed only in the base payment amount,
the majority of workers would choose the HIT with the higher base
payment, resulting in selection bias.  To avoid this, we used the
following method for randomly assigning treatments.  We first posted a
qualification HIT 
in which workers were paid a small amount (\$$0.05$) if they agreed to
receive notifications about our future tasks.  We made it clear that
they were under no obligation to do the future tasks.  We then
randomly assigned the workers who completed this recruitment task to
different treatments.  For each treatment, we posted a separate HIT.
Workers were only qualified
to see and do the HIT corresponding to their assigned treatment.
Finally, we used the notifyWorkers API call to send notification
emails to workers with a link to their assigned HITs (treatments).
While others have suggested using qualifications to filter out workers
for experiments~\cite{CMP14} or recruiting a panel of workers in
advance~\cite{MS12}, we believe that the approach described here is
novel and of independent interest.

We next describe the treatments:

\squishlist
\item 
  \emph{Low Base:} The base payment was \$$0.50$. No opportunity for a
  bonus was given.  This was our control.
\item 
  \emph{High Base:} The base payment was \$$1.50$. No opportunity for a bonus was given.
\item 
  \emph{Unexpected Bonus:} The base payment was \$$0.50$.  After accepting
  the HIT, workers were told they would receive an additional bonus of
  \$$1$.
\item 
  \emph{PBP:} The base payment was \$$0.50$.  In addition to the base
  payment, workers could earn a bonus of \$$1$ if they correctly
  labeled 80\% of the image pairs as identical or not. Workers were
  informed of the bonus and rules for receiving the bonus before
  accepting the HIT.
\squishend

\ignore{
Note that in the \$$0.50$ Base $+$ \$$1.00$ Bonus treatment workers were
informed of the bonus only after accepting the HIT.  This was most likely a
surprise to the workers.  On the other hand, workers in the \$$1.50$ Base
treatment and the \$$0.50$ Base $+$ \$$1.00$ PBP treatment got full
information about the payment structure before accepting the HITs.
}

Note that the payment amounts in the High Base and Unexpected Bonus
treatments are the same.  The difference is only how and when the
payments were described.

\subsection{Results}

\begin{figure}[h]
   \centering
   \includegraphics[width=1\columnwidth, keepaspectratio]{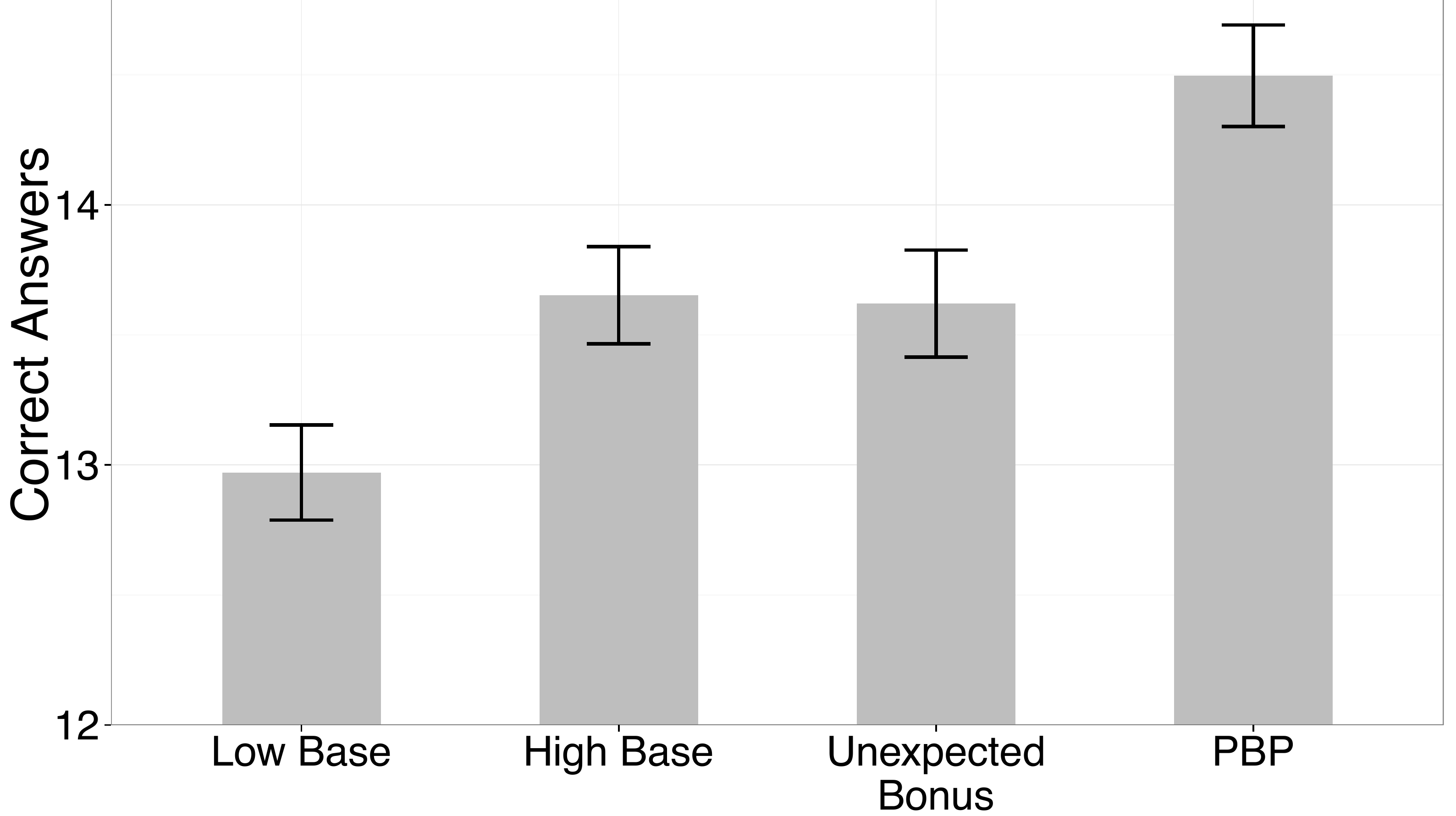}
   \caption{The effect of different payment
     schemes on work quality in the spot the differences task. Error bars
   indicate the mean $\pm$ one standard error.}
   \label{fig:SD_figure}
\end{figure}

We randomly chose $800$ workers from the pool that completed the
qualification HIT and randomly assigned them to the four treatments,
$200$ workers per treatment.  After assigning qualifications
corresponding to each treatment, we posted the HITs for each 
simultaneously and sent each worker a notification with a link to
their treatment's HIT.  We conducted a chi-squared test to check for
significant differences in the number of participants who finished the
four treatments and found none ($p = 0.90$).  In the analysis we make
six comparisons, described below. We did this analysis using an ANOVA
with one-sided, planned comparisons \cite{seltman-book} and report p-values that have been
corrected for these multiple comparisons.  The results are shown
in Figure~\ref{fig:SD_figure}.

%

Similar to the proofreading experiment described in
Section~\ref{sec:whether}, simply paying more resulted in higher
quality work.  The High Base treatment had a significantly higher
number of correct answers than the Low Base treatment $(p=0.030)$.
Similarly, the Unexpected Bonus treatment had a significantly higher
number of correct answers than the Low Base treatment $(p=0.047)$.
Figure~\ref{fig:SD_figure} shows no significant difference between the
High Base and the Unexpected Bonus treatments.  This suggests that
there was no ``unexpected bonus effect'' in contrast
to~\citet{GLM14}\footnote{ Note that our experimental setting is not
  the same as theirs. They ran experiments on
  oDesk for tasks with much longer working hours (3 hours) while our
  experiment is on MTurk and our tasks only last for on average 8.8
  minutes.}
 The absence of any reciprocity
effect due to the unexpected bonus suggests that workers were doing
better work to increase the probability (according to their prior
assumptions) that their work got accepted and thus earn the higher
pay.

We also observe that workers in the PBP treatment outperformed workers
in all other treatments $(p < 0.005)$. This suggests that workers are
rational to some degree and are willing to exert more effort to
increase their chances of receiving higher payments.  Note that in
this experiment workers knew \emph{before} they accepted the HIT that
they could earn a bonus, in contrast to the experiment described in
Section~\ref{sec:whether} in which workers were informed of the
opportunity to earn a bonus only \emph{after} they accepted the HIT.
We have therefore shown that PBPs can work whether or not the
opportunity for a bonus is unexpected.

%

%% file: exp4.tex
We have shown that PBPs incentivize higher quality crowdwork on two
specific tasks, proofreading and spotting differences in images. It is
natural to ask whether our results generalize, and in
particular, what properties of a task open up the possibility of
performance improvements with PBPs.

\citet{CH99} note that in the context of economics lab experiments,
performance-based incentives tend to improve quality for
\emph{effort-responsive} tasks, tasks for which it is possible to
generate higher quality work by exerting additional effort (presumably
without requiring \emph{too much} effort).  One might ask if the same
is true in a crowdsourcing setting. More specifically, we {investigate a hypothesis} that
whether, and to what extent, a task is effort-responsive is an important
reason for whether or not PBPs work for this task. {We find an empirical correlation between the two, which we interpret as a strong evidence in favor of this hypothesis.}  
{Since} it is difficult to directly measure how much effort a worker has put into a task, we use
the time a worker spent on a HIT as a proxy measure for effort, and
examine the relationship between time spent and quality of work.

\begin{figure*}[ht]
   \centering
   \subfigure[Proofreading]{
   	\includegraphics[width=0.45\columnwidth, keepaspectratio]{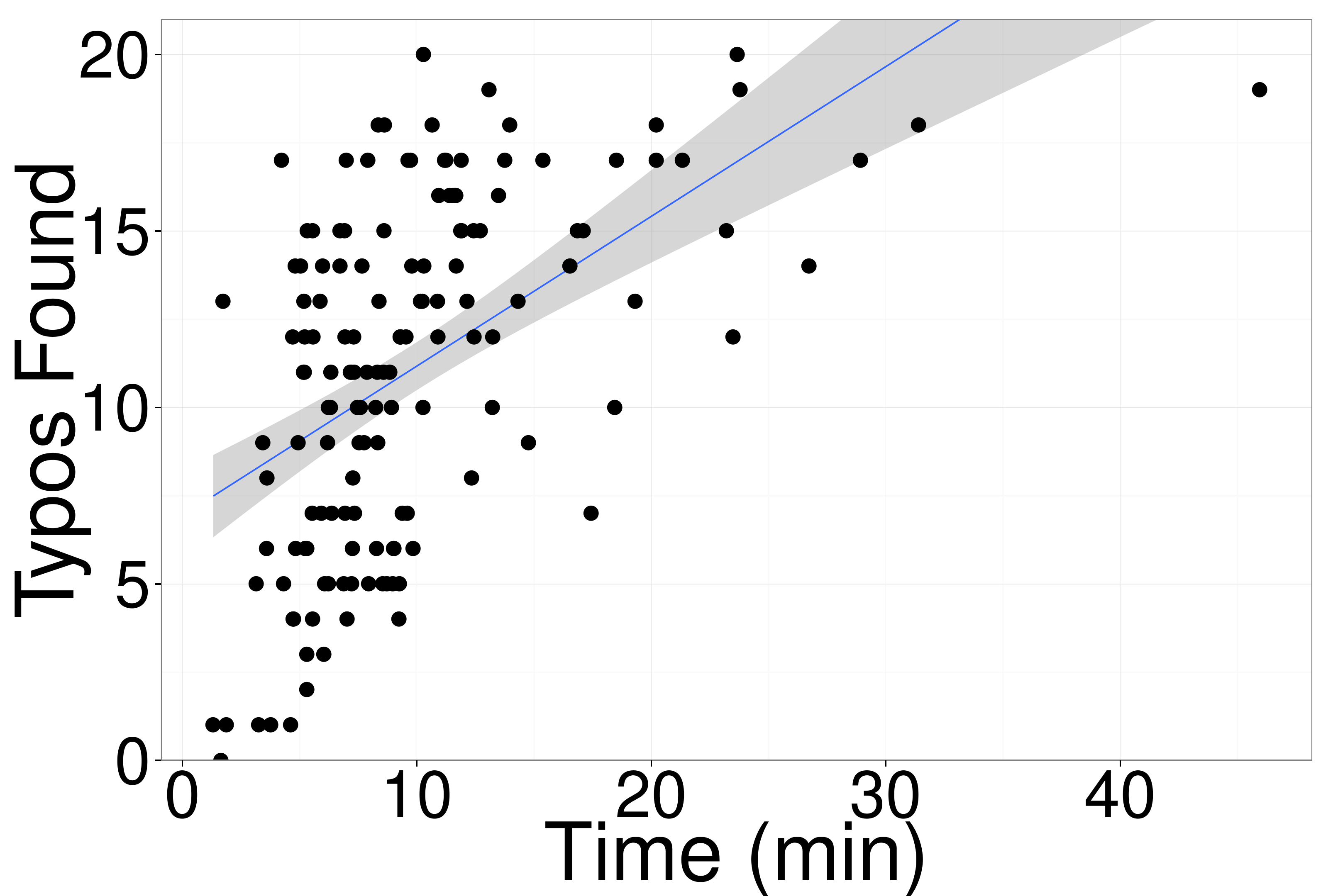}
      \label{fig:proofreading_effort}
   }
   \subfigure[Spot-the-Difference]{
   	\includegraphics[width=0.45\columnwidth, keepaspectratio]{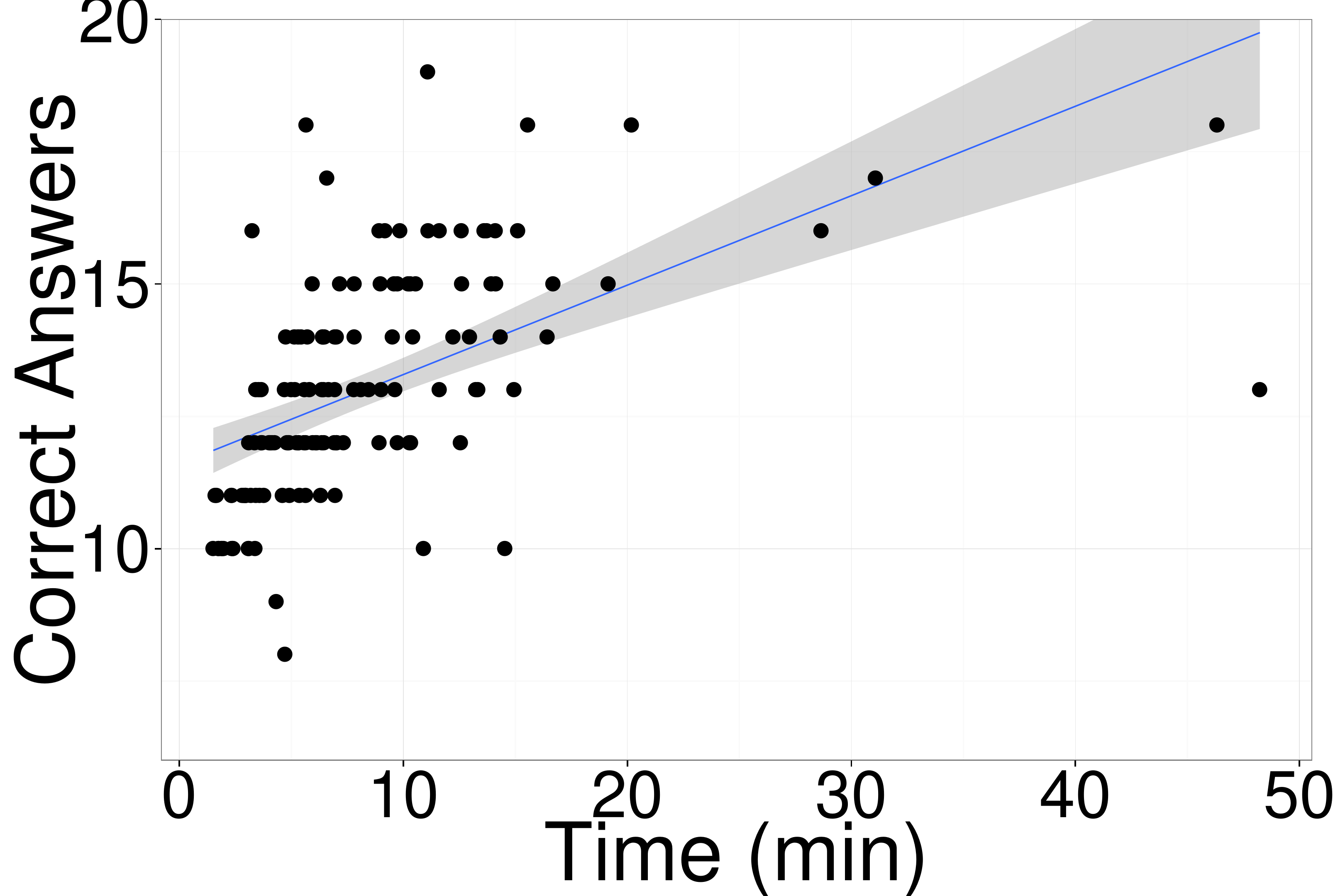}
      \label{fig:SD_effort}
   }
   \subfigure[Handwriting Rec.]{
   	\includegraphics[width=0.45\columnwidth, keepaspectratio]{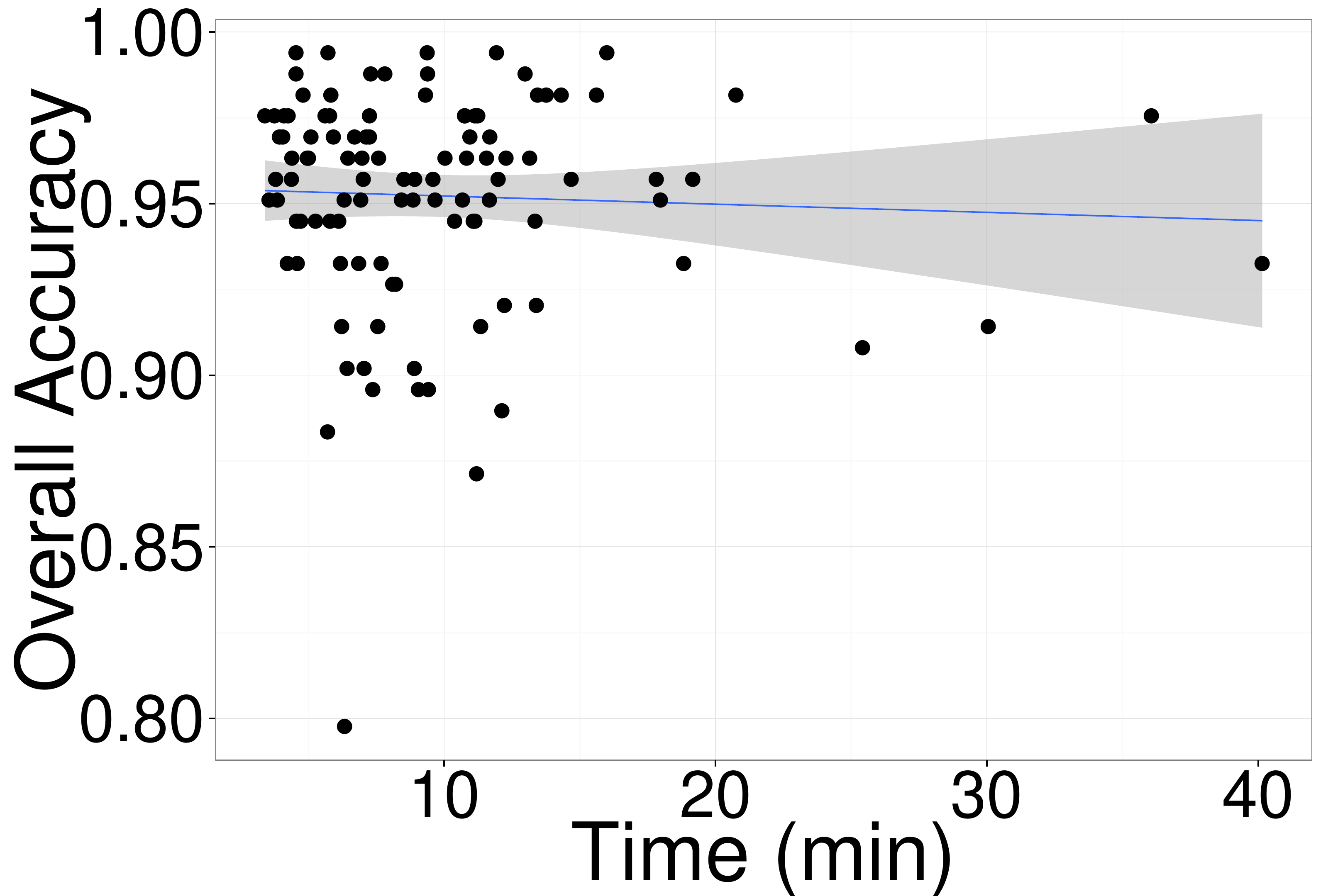}
      \label{fig:handwriting_effort}
   }
   \subfigure[Audio Transcription]{
   	\includegraphics[width=0.45\columnwidth, keepaspectratio]{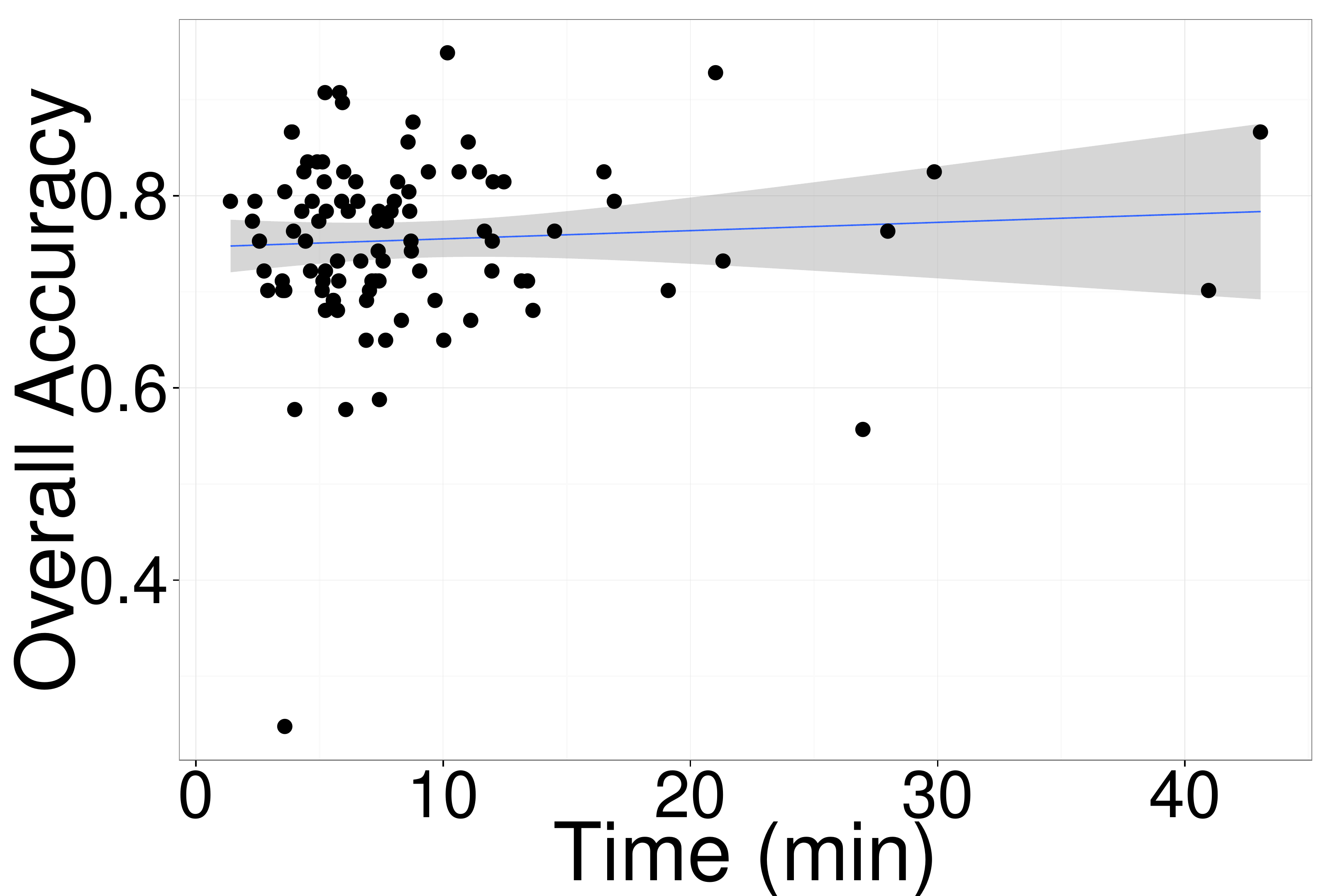}
      \label{fig:audio_effort}
   }
   \caption{Time vs. quality for effort
   responsive tasks in panels \ref{fig:proofreading_effort} and \ref{fig:SD_effort},
   and non-effort responsive tasks in  panels \ref{fig:handwriting_effort}
   and \ref{fig:audio_effort}.  The blue lines indicate the regression line
   and the shaded areas represent the 95\% confidence interval around it.
   Results are similar when outliers are excluded from the analysis.}
	\label{fig:effort_based}
\end{figure*}

Figures~\ref{fig:proofreading_effort} and~\ref{fig:SD_effort}
illustrate this correlation for the proofreading and spot-the-difference
tasks respectively.  Each shows the amount of time that a worker spent
on the HIT versus the quality of his work (measured as number of typos
found or number of correctly labeled image pairs as before).
For the proofreading task, regressing the number of correct typos found on
the amount of time a worker spent shows that every minute is correlated
with finding another 0.42 typos on average ($p<0.001$).  Similarly, for the
spot-the-difference task, regressing the number of pairs correctly
identified on the amount of time a worker spent shows that every minute is
correlated with another 0.17 correct answers ($p<0.001$).
We see
that, in general, workers who spent more time on our tasks generated
better quality work. We observe similar trends in all treatments, but
include only workers in the control groups in the plots since they are
most comparable across tasks.  This is evidence that the tasks on
which we observed improvements from PBPs are effort-responsive.

To further explore this {hypothesis} and the generalizability of our
results, we examined the effects of PBPs on two additional tasks,
handwriting recognition and audio transcription.  For consistency with
our previous experiments, we maintained a base payment of \$$0.50$ (adjusting the task lengths to maintain an hourly rate of roughly $\$6$) and
a bonus of \$$1$ when applicable.  Workers were randomly assigned
to treatments after accepting the HIT in the same way as in the experiments described
in Sections~\ref{sec:whether} and~\ref{sec:how}.

\subsection{Handwriting Recognition: Design}

For the handwriting recognition task, workers were shown two images
containing handwritten text and asked to transcribe the text.  The
images were collected from the IAM Handwriting
Database~\cite{MB02}.
One contained 89 words and the other 74. As in our other experiments,
workers in the control treatment received only the base payment. We
also used a PBP treatment in which workers were told that for one of
the images, their transcription would be compared against a gold
standard solution and they would receive the bonus if they correctly
transcribed 90\% of the words in the image.  They were not told which
image would be used to assess their accuracy.  This payment rule was
used to make the task appear more like a realistic MTurk task.

\subsection{Handwriting Recognition: Results}

Data was collected from 220 workers.  As shown in
Figure~\ref{fig:handwriting_effort}, the quality of work produced was
not significantly correlated with the time a worker spent on the task.
In other words, this task does not appear to be effort-responsive.
Moreover, we did not find a significant difference between the
accuracy of workers in the control group versus the PBP treatment via
a one-sided t-test.
 We note that one of the two images given to
workers was chosen because it was especially difficult for workers in
a pilot study.  Restricting our analysis to just this image did show a
marginal effect, with workers in the control group averaging 92.3\%
accuracy and workers in the PBP treatment averaging 93.5\%
($p=0.055$).  Since PBPs only had a small, marginal effect on even the
most difficult handwriting recognition image, we conclude that PBPs
only have a small effect on this task in general.

One reason why PBPs did not have a strong effect on quality for this
task may be that there was a ceiling effect, as discussed in the context of
incentives in lab experiments by~\citet{CH99}.  As
Figure~\ref{fig:handwriting_effort} shows, the average accuracy over
the two articles of the workers in the control group was 95.2\%,
leaving little room for PBPs to have impact. A related explanation
is that most of handwritten words in our data sets were trivial to
recognize. Over 80\% of words were correctly transcribed by over 90\%
of workers.

It is possible that PBPs would have improved performance
if we had chosen a different threshold.  However, since the average
performance was already very high, there was little room for
experimentation.  If a better threshold exists, it would be difficult
for a requester to identify.

\vspace{0.5cm}
\subsection{Audio Transcription: Design}

For the audio transcription task, workers were asked to transcribe 10
audio clips, each of which contained approximately 5 seconds of
speech.  The audio clips were collected from
VoxForge\footnote{http://www.voxforge.org/} and we intentionally
chose clips from speakers with heavy accents to increase the
difficulty of the task and avoid the ceiling effect.
Once again workers in the control treatment
received only the base payment.  We additionally included three PBP
treatments with different thresholds of quality required to receive
the bonus.  In these treatments, workers were told that their answers
for 5 of the 10 clips would be compared against gold standard answers,
and that they would receive the bonus if they correctly transcribed
80\%, 85\%, or 90\% of the words respectively.  These threshold values
were chosen based on quality observed in a pilot experiment on this
task in order to cover the range of thresholds that we believed would
be most likely to lead to quality improvements with PBPs.  We note
that audio transcription is one of the most common tasks on MTurk.

\subsection{Audio Transcription: Results}

We collected data from 400 workers.  As Figure~\ref{fig:audio_effort}
shows, quality and time are not significantly correlated. That is, audio
transcription does not appear to be effort-responsive. Furthermore, we did not find a significant difference between any of the three PBP
treatments and the control group.
Since the average accuracy in the
control group was only 75.4\%, this cannot be fully
explained by a ceiling effect as may have been the case for
handwriting recognition. One might ask if there are certain hard words
for which PBPs did improve performance.  We took a closer look at the
data and found that this does not appear to be the case.  Of the 97 words included in the 10 clips, workers'
combined accuracy was better in the 85\% threshold PBP treatment
compared with the control on 52 words, and better in the control than
the 85\% threshold PBP treatment on 44; essentially these
differences appear to be mostly due to noise.  We observe a similar
pattern for the other PBP treatments.  Additionally, for more than 80\% of the
words, the percentage of workers correctly transcribing the word when
PBPs were offered is close (within $\pm5\%$) to the percentage in the control group. This
could suggest that workers' performance is limited by their
abilities and cannot be improved through PBPs.

\subsection{A Practical Recommendation}

While the results in this section are not causal, they are in line with the {hypothesis that the extent to which a task is effort-responsive is an important reason for whether or not PBPs help improve quality for this task. This} suggests an approach that requesters can use when deciding whether or not to employ PBPs in their own HIT. A requester could run a pilot of their HIT with a small number of workers and a fixed (not performance-based) payment and plot the time that workers spend on the task versus the quality of their work to determine whether {and to what extent} the task is effort-responsive. {A requester may be able to incentivize higher quality using PBPs only if the task is (sufficiently) effort-responsive.}
In this case, the requester must determine whether the boost in quality is worth the extra cost of PBPs.



%% file: model.tex
The worker model in the well-known \emph{principal-agent}
framework \cite{LM02} assumes that a worker chooses a level of effort or
quality to maximize his expected utility, which is simply the expected
payment he receives minus the expected cost of doing the work at the chosen
effort level.  In this section, we show that incorporating the worker's
subjective beliefs about how much quality is required to earn the base and
bonus payments into the worker model allows us to explain the main
observations from our experiments in a parsimonious way. Furthermore, it is
not clear how one might capture our experimental results
using the traditional worker model \emph{without} incorporating the worker's
subjective beliefs. Our model can be
used to reason about the possible consequences of using performance-based
payments.  

%

We assume that a worker views a task and chooses to produce work at a particular quality level in order to maximize his expected utility, defined as his \emph{perceived} expected payment minus the perceived cost of doing the task at a given quality level. In our model, when workers are offered performance-based payments with base payment $p$ and bonus payment $b$, the worker's expected utility is
\begin{align}
\Upbp(q) = p \prbase{q} + b \prbonus{q} - c(q),
\label{eqn:upbp}
\end{align}
where $q$ is the quality level, $\prbase{q}$ and $\prbonus{q}$ denote the
worker's perceived probabilities of receiving the base and
bonus payments with work of quality $q$, and $c(q)$ is the worker's
perceived cost of producing quality $q$.
We posit that $q$ comes from a totally ordered set, and assume that
$\prbase{q}$ and $\prbonus{q}$ are non-decreasing in $q$.

While we typically think of the cost of producing work as positive, capturing the effort the worker must exert to produce work of the chosen quality level, it could in some cases be negative, capturing the subjective intrinsic utility the worker receives from his enjoyment of the task or his satisfaction from a job well done.  To allow for such effects, we make no assumptions on the monotonicity of $c(q)$.

We make a minor \emph{consistent tie-breaking} assumption.
If multiple quality levels maximize the expected utility for
a pair of base and bonus payments, we assume the tie is broken consistently
in the sense that the worker chooses the same quality level
for any payments leading to this particular tie.

\ignore{
For stronger results, one can also assume what we call a \emph{well-behaved model}. Here $q$ takes values on some  interval
    $[q_{\min}, q_{\max}]$
such that $c(q_{\max})>b+p$, and
    $\prbase{q}$, $\prbonus{q}$, and $c(q)$
are differentiable on this interval, and we assume that the subjective probabilities $\prbase{q}$ and $\prbonus{q}$ are strictly increasing in $q$.  Most of our results do not rely on these extra assumptions; we make them only where explicitly stated.
}

\subsection{Consequences of the Worker Model}

Given the worker model, we are able to provide a coherent explanation
of our key observations, including some that are not explained by the standard principal-agent model.
To explain the key observations from our experiments we
compare \eqref{eqn:upbp}, which describes the worker's
utility when there are both a base payment and a bonus payment, with the
utility of a worker with subjective beliefs under the ``standard'' payment
scheme in which workers are offered a base payment only.
The standard payment scheme was used in the control group in our experiments.
In this case, the worker's expected utility is simply
\begin{equation}
\Ustand(q) = p \prbase{q} - c(q).
\label{eqn:ustand}
\end{equation}
\ignore{
When performance-based payments are offered, the worker's expected utility becomes
\begin{equation}
\Upbp(q) = p \prbase{q} + b \prbonus{q} - c(q).
\label{eqn:upbp}
\end{equation}
}
Let $q_\stand$ be the quality level chosen by the worker under this utility function,
 so
    $q_\stand \in \arg\max_q \Ustand(q)$,
and $q_\pbp$ be the quality chosen under \eqref{eqn:upbp}, so $q_\pbp \in \arg\max_q \Upbp(q)$.

\xhdr{Subjective beliefs about acceptance criteria increase quality}
In Section~\ref{sec:whether}, we experimentally showed that the quality of
work produced is higher when workers have subjective beliefs about
acceptance criteria than when the base payment is {explicitly} guaranteed.
We called this an \emph{implicit PBP}.  It is easy to explain in our model.

Consider the standard setting (no
PBPs). With a guaranteed base payment, the worker's utility becomes $p -
c(q)$. Let $q^*$ be a maximizer of this expression. It follows that $q^*$
is a minimizer of $c(q)$, thus  $c(q^*) \leq c(q)$ for all $q$.
For any $q<q^*$, we have $p\prbase{q} \leq p\prbase{q^*}$ given the monotonicity
of $\prbase{q}$. Therefore, for any $q<q^*$, $\Ustand(q) \leq \Ustand(q^*)$.
Given the consistent tie-breaking assumption, the worker will not choose to generate work of quality $q$ if $q<q^*$. Therefore, if $q_\stand$ is the maximizer of \eqref{eqn:ustand} then $q_\stand \geq q^*$.  Thus $q_\stand$, the optimal quality level with no guaranteed base payment, is greater than or equal to $q^*$, the optimal quality level with guaranteed base payment, which we observed experimentally in Figure~\ref{fig:qcp_work}.  It is possible to strengthen this result to strict inequality ($q_\stand > q^*$) with a few additional assumptions.%
\footnote{\label{fn:well-behaved}%
{In multiple places in this section, to make inequalities on quality strict, we could first assume that $q$ takes values on some interval $[q_{\min}, q_{\max}]$ such that $c(q_{\max})>b+p$, and that $\prbase{q}$, $\prbonus{q}$, and $c(q)$ are differentiable on this interval.  We must then also assume that the subjective probabilities $\prbase{q}$ and $\prbonus{q}$ are strictly increasing in $q$. In this particular instance, under these assumptions, since $c'(q^*)=0$, we have $\Ustand'(q^*)>0$, which implies that $q^*$ cannot be the maximizer of $\Ustand'(q^*)$, so $q_\stand > q^*$.}}
%
\ignore{
Further, if the model is well-behaved, then $q_\stand > q^*$. {Indeed, since $c'(q^*)=0$ and $\prbase{q}$ is strictly increasing, we have $\Ustand'(q^*)>0$, which implies that $q^*$ cannot be the maximizer of $\Ustand'(q^*)$. So it must be the case that $q_\stand > q^*$.}
}
Similar arguments
can be made for the setting with PBPs.


\xhdr{Higher payments increase quality}
The experiments in Sections~\ref{sec:whether} and~\ref{sec:how} demonstrated that
increasing the base payment (or the unconditional bonus) can increase quality. This is also easy to explain in our model. Consider increasing the base payment in the standard setting (no PBPs).
Let $q_\delta$ be the maximizer of
    $U_\delta(q) = (p + \delta) \prbase{q} - c(q),$
the utility of the worker if the base payment were $p + \delta$ instead of
$p$, where $\delta > 0$.
By optimality of $q_\stand$, for all $q$,
$$ p\prbase{q} - c(q) \leq p\prbase{q_\stand} - c(q_\stand). $$
Since $\prbase{q}$ is non-decreasing in $q$, $q < q_\stand$ implies
$\delta\prbase{q} \leq \delta\prbase{q_\stand}$.
Combining the last two inequalities,
$U_\delta(q) \leq U_\delta(q_\stand)$ for all $q<q_\stand$.  Therefore, given the consistent tie-breaking assumption, $q_\delta$, the optimal quality level under higher pay, is greater than or equal to $q_\stand$, the optimal quality level under the standard setting. Again, strict inequality is achievable with the {additional assumptions from Footnote~\ref{fn:well-behaved}}.
%
\ignore{
Further, if the model is well-behaved, then $q_\stand > q_\delta$. This is because
    $U_\stand'(q_\stand)=0$,
and consequently $U'_\delta(q_\stand)>0$.
}
Similar arguments can be made for increasing the base payment with PBPs, and for increasing the bonus payment.

{These conclusions hold only when uncertainty about receiving the payment is included. They would not hold in the standard principal-agent model, where increasing a guaranteed payment does not increase quality.}

\xhdr{Performance-based payments (significantly) increase quality}
A key result from our experiments in Sections~\ref{sec:whether}-\ref{sec:why}
is that PBPs can, in fact, increase quality in a significant way.  This can be explained
from our model too.

We have that the quality improves ($q_\pbp \geq q_\stand$, or $q_\pbp > q_\stand$ with additional assumptions)
%
\ignore{ (with a strict
inequality assuming the well-behaved model)
}
as a special case of the
argument above. However, this statement is relatively weak, as it does not
say anything about the magnitude of the improvement, i.e., the difference
$q_\pbp - q_\stand$.
We would like this difference to be large; a
requester might not want to take on the extra costs of PBPs if the
increase in quality is small.  In order to gain some intuition for
when it is or is not possible to obtain a sufficiently large
improvement in quality, we consider several special cases as examples.

As a first example, consider the case in which the worker does not
have fine-grained control over the precise quality of his work, but
can only choose between two options: high-quality, denoted $q_\high$,
or low quality, $q_\low$.
{Then PBPs work if and only if the worker's optimal quality
level is $q_\low$ without PBPs, and $q_\high$ with PBPs. In formulas,
    $\Ustand(q_\high) < \Ustand(q_\low)$ and $\Upbp(q_\high) > \Upbp(q_\low)$},
or
\begin{align*}
& p \left(\prbase{q_\high} - \prbase{q_\low}\right) < c(q_\high) - c(q_\low) \\
&\qquad < p \left(\prbase{q_\high} - \prbase{q_\low}\right) + \\
&\qquad \;\;\;\;\; b\left(\prbonus{q_\high} - \prbonus{q_\low}\right) .
\end{align*}
In words, the extra cost to produce high-quality work must be less than the extra benefit the worker would receive in terms of expected payments if the bonus is included, and bigger than the extra benefit he would receive with standard payments. Examining this expression gives us intuition about when we might expect PBPs to help. First, $c(q_\high) -c(q_\low)$ cannot be too large. It must be possible for the worker to substantially increase his quality with additional effort (and at a reasonable cost)---essentially, the task must be effort-responsive, as conjectured in Section~\ref{sec:generalizability}.  Second, $c(q_\high) -c(q_\low)$ cannot be too small (in particular, compared with $\prbase{q_\high} - \prbase{q_\low}$ and $p$) or PBPs are unnecessary to achieve high quality.  This is a partial explanation for why we did not see improvements from PBPs in the handwriting recognition task when the cost of producing high quality was already small.  Third, the bonus $b$ must be set large enough.  This could partially explain our observation that PBPs did not help in the proofreading task with a very small bonus.  Finally, the difference $\prbonus{q_\high} - \prbonus{q_\low}$ must be high enough.  This could explain why we observed that PBPs did not help when the threshold for receiving a bonus is set too low.


As a second tractable example, suppose that the worker has
fine-grained control over quality, but has no uncertainty over whether
the bonus will be obtained, i.e., 
\begin{align*}
	\prbonus{q} =
    \begin{cases}
    	1 & \text{if $q\geq\bar{q}$}\\
        0 & \text{otherwise}
    \end{cases}
\end{align*}
for some threshold value $\bar{q}$.
It is easy to show that either
$q_\pbp = q_\stand$ (if the worker prefers to do less work and pass up the bonus)
or $q_\pbp \geq \bar{q}$ (if the worker prefers to do more work to receive the bonus). PBPs are useful if and only if
$\bar{q}$ is sufficiently higher than $q_\stand$ (according to the needs of the requester),
and for some $q \geq \bar{q}$,
    $\Upbp(q) > \Upbp(q_\stand)$,
or equivalently
\begin{align}\label{eq:model-example-Qbar}
c(q) - c(q_\stand) <  p(\prbase{q} - \prbase{q_\stand}) + b.
\end{align}
Again we see that for PBPs to help it must be possible for the worker
to increase his quality with additional effort at a reasonable cost, and the bonus must be set
sufficiently large.

\eqref{eq:model-example-Qbar} provides a concrete and simple way to think about whether it would help to increase the bonus threshold $\bar{q}$. For fixed payments $p$ and $b$, it is optimal to choose as $\bar{q}$ the largest $q$ which satisfies \eqref{eq:model-example-Qbar}.
However, $\bar{q}$ is a \emph{perceived} threshold and cannot always be controlled directly.  Instead, it may help to alter workers' perception, perhaps without even changing the objective bonus rule. For example, in the proofreading experiment from Section~\ref{sec:how} the ``5 typos" bonus rule is roughly equivalent to the ``25\%" bonus rule in terms of when bonus payments are awarded, yet workers react differently. One possible explanation for this general phenomenon is a difference in the perceived value $\bar{q}$.

\subsection{Comparison with Principal-Agent Model}

In the standard principal-agent model, a worker maximizes his payment minus the intrinsic cost of his effort.
We deviate from the standard model in that we include the worker's subjective beliefs about how high quality his work must be to be paid. In particular, while the objective probability of receiving the base payment is often 1, the subjective probability may be much smaller, depending on the quality level. This feature allows us to capture some effects that are not captured by the standard model, e.g., that increasing the base payment may increase the quality of work that the worker chooses to produce, while removing uncertainty about the base payment may decrease quality.

Our model may be useful as a more realistic foundation for theoretical work.
As an example, consider \citet{HSV14}, a recent theoretical paper on
the optimization of
PBPs. While that paper posits the standard principal-agent model, all
results carry over to our model.  (A proof of this fact is omitted due to space constraints.)

\ignore{ 
\footnote{A proof sketch for this statement is as follows. \citet{HSV14} observe that all of their results work under a mild condition on worker behavior. Translating to our model, this condition states that increasing the bonus payment cannot decrease the objective probability of the bonus being awarded. It is easy to see that the condition is satisfied. As we've seen above, increasing the bonus payment can only increase the resulting quality level chosen by the worker, which in turn can only increase the distribution of the quality measure observed by the requester (in the sense of first-order stochastic dominance), and increasing the quality measure can only increase the objective probability of assigning the bonus. The middle step in this argument is a very reasonable assumption on the observable quality measure as a randomized function of the chosen quality level, and the last step is a very reasonable property of the rule for assigning the bonus depending on the observable quality measure.}
} 

%% file: conclusion.tex

We describe the results of a series of experiments studying the effect of performance-based payments on the quality of crowdwork on Amazon Mechanical Turk.  Our goal is to identify properties of the payment, payment structure, and task that allow for quality improvements using PBPs.

We find that PBPs can improve the quality of submitted work for some tasks, but are not likely to for others. We identify the extent to which a task is \emph{effort-responsive} as a potential important reason for whether or not PBPs work for this task. This leads to an actionable insight for requesters. When considering whether or not to use PBPs, one could first run a pilot experiment to determine whether a task is effort-responsive by examining the correlation between time spent and quality.  If additional time spent leads to a sufficiently high boost in quality, we would expect PBPs to improve performance.


We find strong evidence for what we call the \emph{implicit PBP effect}:
workers may have their own subjective beliefs about the quality of work
they must submit to have their work accepted, which makes them view fixed
payments as implicitly performance-based.  Workers may also have subjective
beliefs about the likelihood of receiving the bonus when payments are
explicitly performance-based.  This should be taken into account when
designing a payment scheme. For example, we find that in some cases a
requester can incentivize higher quality work by defining the bonus threshold
relative to other workers or with respect to gold standard data
that the workers do not have access to.

We show that in order for PBPs to improve quality, the bonus payment offered must be sufficiently large, but that there are diminishing returns for further increasing this payment.
This partially explains existing negative results on the effectiveness of PBPs in crowdsourcing markets.  We provide evidence that when PBPs improve quality, they do so for a wide range of quality thresholds.

Finally, we suggest a theoretical model of workers' behavior that captures all of the above effects and explains several outcomes we observed in our experiments. We believe this model may be useful in further work on crowdsourcing markets, both as a concrete way to think about the consequences of using this or that payment scheme in practice, and as a more realistic foundation for theoretical work compared to the standard principal-agent model.

%% file: ack.tex
This research was partially supported by the National
Science Foundation under grant IIS-1054911.